\documentclass[aps,prb,twocolumn,preprintnumbers,amsmath,amssymb,showpacs]{revtex4}
\usepackage{bbm}
\usepackage{cases}
\usepackage{mathbbol}  
\usepackage{graphicx}
\usepackage{graphicx}
\usepackage{dcolumn}
\usepackage{bm}
\usepackage{mathrsfs}
\usepackage{amssymb}
\usepackage{txfonts}
\usepackage{epsfig,amsmath,amssymb,color}
\begin{document}
\title{Dynamical Mean Field Theory for the Bose-Hubbard Model}

\author{Wen-Jun Hu}
 \affiliation{Department of Physics, Renmin University of China, Beijing 100872, People's Republic of China}%

\author{Ning-Hua Tong}%
\email{nhtong@ruc.edu.cn}
 \affiliation{Department of Physics, Renmin University of China, Beijing 100872, People's Republic of China}%

\begin{abstract}

The dynamical mean field theory (DMFT), which is successful in the
study of strongly correlated fermions, was recently extended to
boson systems [Phys. Rev. B {\textbf 77 }, 235106 (2008)]. In this
paper, we employ the bosonic DMFT to study the Bose-Hubbard model
which describes on-site interacting bosons in a lattice. Using exact
diagonalization as the impurity solver, we get the DMFT solutions
for the Green's function, the occupation density, as well as the
condensate fraction on a Bethe lattice. Various phases are
identified: the Mott insulator, the Bose-Einstein condensed (BEC)
phase, and the normal phase. At finite temperatures, we obtain the
crossover between the Mott-like regime and the normal phase, as well
as the BEC-to-normal phase transition. Phase diagrams on the
$\mu/U-\tilde{t}/U$ plane and on the $T/U-\tilde{t}/U$ plane are
produced ($\tilde{t}$ is the scaled hopping amplitude). We compare
our results with the previous ones, and discuss the implication of
these results to experiments.

\end{abstract}

\maketitle
\section{introduction}
The ultracold atoms trapped in an optical lattice have aroused
growing interests in recent years. By regulating the various
parameters of the standing wave laser fields that create the optical
potentials, such as the laser power and wave length, many
theoretical models in the condensed matter physics can be realized
experimentally, especially those for the strongly correlated many
body systems.\cite{JZ} In particular, bosons in a lattice have been
widely studied in theory and experiment. The investigation of the
correlated bosons can be traced back to the study of
$^4$He.\cite{MM} Recently D. Jaksch has pointed out that the
Bose-Hubbard model (BHM) Eq.(\ref{eq:1}) can well describe the
ultracold boson atoms in an optical lattice, if one assumes a
short-range pseudo potential interaction between the atoms and that
the Wannier functions are well localized on the lattice
site.\cite{JBZ,BDZ}

The BHM has been studied using various analytical and numerical
methods. In their seminal paper, M. P. A. Fisher \emph{et al.} used
field theoretical approach to investigate the ground state of this
model and obtained the superfluid (SF)-Mott insulator (MI)
transition on the mean field level.\cite{FF} The Mott insulator is
an incompressible state where integer number of bosons are localized
on each site, while the superfluid phase is compressible and has
nonlocal boson wave functions. As an interesting phase, the ground
state of MI is considered as a good candidate to realize the qubits
for quantum information processing.\cite{JZ,MB,MZ} For weakly
interacting bosons for which the fluctuations around the mean-field
state are small, the Bogolubov theory or the Gross-Pitaevskii
equation applies but both fail to predict the SF-MI
transition.\cite{OS,ST} Beyond the mean field level, methods that
can tackle the strong correlations have been applied
\cite{FF,OS,JBZ,SR,PF,BUV}, including the Gutzwiller
approach\cite{RK,KB}, Bethe ansatz\cite{K}, time-dependent
variational principle method\cite{AP}, slave boson
approach\cite{DS,YC,MFZ}, the strong coupling
expansion\cite{FM,FMb,EM,SD}, variational method based on mean field
theory\cite{SAP}, and the effective action approach\cite{SAP,BP},
{\it etc}. Recently the cavity method based on the Bethe lattice is
also applied to BHM\cite{SZ}. The numerical tools such as quantum
Monte Carlo\cite{BZ,WB,PP,CS,KT,CSS,CSSS} and density matrix
renormalization group\cite{PR,KM} are used frequently for the
unbiased studies.

The physics of the BHM depends critically on its spatial dimension. For a one dimensional
system, the Mermin-Wagner theorem\cite{MW} excludes the possibility of Bose-Einstein
Condensation (BEC) at finite temperatures. In the strong interaction regime, the bosons behave
as the Tonks-Girardeau gas whose properties are similar to the noninteracting
fermions.\cite{AG,BDZ,WT} The BEC in two dimensions can be viewed as a quasi-condensate, and
the Mott transition is of the Kosterlitz-Thouless type.\cite{BDZ} The MI is strictly defined
at zero temperature and the MI-BEC transition is a quantum transition well defined at zero
temperature. However, there is a finite temperature range where the occupation is fixed at an
integer. As temperature increases, the system changes from this Mott-like regime into the
normal phase through a smooth crossover. For $D>2$, the SF phase also transits into the normal
phase at a finite critical temperature.\cite{G,KT}

Experimentally, the BHM has been realized in the system of alkali metal atoms in an optical
lattice. The SF to MI transition has been observed in $1D$, $2D$ and $3D$, by changing the
depth of the optical lattices.\cite{SE,SP,GB} These studies focused on the MI and the SF phase
mainly in $2D$ and $3D$\cite{G,SP,GBc,FB,SPa} at finite temperatures. The temperature in these
studies is a key factor and its effects on the observation cannot be ignored. Recent studies
show that due to the finite temperature effects, the sharp peaks in the momentum distribution
commonly adopted to identify the SF cannot be used as a reliable signature.\cite{Gr,DH,KT} Due
to technical difficulties, accurate studies for high dimensional system ($D>2$) at finite
temperatures that match the experiments are still highly desirable.

Here, we are interested in the strongly correlated bosons on a high
dimensional lattice ($D>2$) at finite temperatures, which received
less attention in previous theoretical studies. One suitable method
for our purpose is the dynamical mean field theory (DMFT). DMFT is
an exact theory in infinite spatial dimensions.\cite{MV} As an
approximation for finite dimensional systems, it has been widely
used in the study of strongly correlated fermion systems and
received much success.\cite{GK} In a recent work, K. Byczuk and D.
Vollhardt extended the idea of DMFT to correlated boson systems and
applied it to the bosonic Falicov-Kimball model.\cite{BV} In their
theory, instead of the usual way of scaling the hopping amplitude in
the fermionic DMFT, they used a different scaling ansatz for bosons:
scaling $t_{ij}\longrightarrow \tilde{t}_{ij}/z^{|i-j|}$ for terms
containing the anomalous averages $\langle b\rangle$ or $\langle
b^{\dag}\rangle$, and scaling $t_{ij}\longrightarrow
\tilde{t}_{ij}/\sqrt{z}^{|i-j|}$ for others. Such a new scaling is
used to guarantee that the energy density is finite as the spatial
dimension goes to infinity, even if anomalous averages are involved
in the boson systems. It should be noted that the derivation of the
above B-DMFT equations is not unique. Recently, the essentially
identical equations are also obtained by using an uniform scaling of
$t \rightarrow \tilde{t}/z$ and keeping up to the subleading order
in the $1/z$ expansion.\cite{HH} In this paper, we apply B-DMFT to
the BHM. The resulting effective bosonic impurity model is solved by
exact diagonalization (ED) method\cite{LB,BV}. Results are presented
for various phases at finite temperatures and compared to other
theories and the experiments.

This paper is organized as follows. In Sec. II, we briefly introduce the single band Bose
Hubbard model. We present the B-DMFT equations for the BHM, detail the impurity solver that we
use, and give tests and benchmarks. In Sec. III, the main results of B-DMFT are shown and
discussed. Sec. IV is a conclusion. We put some technical details in Appendices.

\section{model and method}
\subsection{Bose Hubbard Model}
The single band BHM is defined by the Hamiltonian below
\begin{eqnarray}\label{eq:1}
H=-\sum_{\langle i,j\rangle}t_{ij}b_{i}^{\dagger}b_{j}
  +\frac{U}{2}\sum_{i}n_{i}(n_{i}-1)-\mu\sum_{i}n_{i},
\end{eqnarray}
where $b_{i}^{\dagger}$ and $b_{i}$ are the boson creation and annihilation operators on site
$i$, respectively. They obey the commutation relation $[b_{i},b_{j}^{\dagger}]=\delta_{i,j}$.
$n_{i}=b_{i}^{\dagger}b_{i}$ is the boson number operator on the site $i$. Here, we consider
the hopping amplitude $t_{ij}=t$ which is nonzero only for the nearest neighbors and $U$ is
the on-site energy. Feshbach resonances can be used to change the interaction strength over a
wide range, even from repulsive to attractive.\cite{BDZ} In this paper we study the repulsive
BHM with $U\geq0$. We add the chemical potential $\mu$ which controls the number of bosons in
the grand canonical ensemble. For simplicity, we take the density of states of the Bethe
lattice, which is semicircular in the limit of infinite coordinations,
\begin{eqnarray}\label{eq:2}
 D(\epsilon)=\frac{1}{2\pi t^{2}}\sqrt{4t^{2}-\epsilon^{2}}, &(|\epsilon|\leq 2t).
\end{eqnarray}

\subsection{B-DMFT Equations}
In DMFT, a lattice model is mapped into a single impurity problem
with the self-consistently determined bath spectra. It becomes exact
when the spatial dimension is infinite and hence ignores the spatial
fluctuations from the outset. However, it fully takes into account
the temporal fluctuations (imaginary time).\cite{GK} The key
ingredient to extend the DMFT to bosons is a proper scaling of the
hopping amplitude of bosons in the limit of infinite
dimensions.\cite{BV}

We adopt the ansatz of scaling in Ref.~\onlinecite{BV} and implement similar derivations for
the BHM. The detail of derivations can be found in Ref.~\onlinecite{BV}. Here we present only
the final B-DMFT equations. For simplicity, here we use the Nambu representation\cite{N} for
the boson operators
$\mathbf{b}^{\dag}(\tau) \equiv \left(\begin{array}{cc}
                         b^{\dag}(\tau) & b(\tau) \\
                         \end{array}\right)$,
and for the on-site interacting Green's functions (GFs) as in
Ref.~\onlinecite{BV}
\begin{equation} \label{eq:3}
\begin{split}
\mathbf{G}(\tau-\tau')&\equiv -\langle T_{\tau}[\mathbf{b}(\tau)\mathbf{b}^{\dag}(\tau')]\rangle \\
                      &=\left(\begin{array}{cc}
                        -\langle T_{\tau}[b(\tau)b^{\dag}(\tau')]\rangle
                      & -\langle T_{\tau}[b(\tau)b(\tau')]\rangle \\
                        -\langle T_{\tau}[b^{\dag}(\tau)b^{\dag}(\tau')]\rangle
                      & -\langle T_{\tau}[b^{\dag}(\tau)b(\tau')]\rangle \\
                        \end{array}
                        \right) \\
                      &=\left(\begin{array}{cc}
                        G_{1}(\tau-\tau') & G_{2}(\tau-\tau') \\
                        G_{3}(\tau-\tau') & G_{4}(\tau-\tau') \\
                        \end{array}
                        \right).
                        \end{split}
\end{equation}
According to the definition, the following relations hold for the components
$G_{3}(\tau-\tau')=G_{2}^{\ast}(\tau-\tau')$ and $G_{4}(\tau-\tau')=G_{1}^{\ast}(\tau-\tau')$.

The action for the effective impurity model obtained through the cavity method \cite{GK}
reads,
\begin{eqnarray}\label{eq:4}
S_{eff} &= \int_{0}^{\beta}\mathrm{d}\tau
       \int_{0}^{\beta}\mathrm{d}\tau'
         \mathbf{b}_{0}^{\dag}(\tau)
         \left[-\mathbf{\mathcal{G}}_{0}^{-1}(\tau-\tau') \right]
         \mathbf{b}_{0}(\tau') \nonumber \\
       & +\int_{0}^{\beta}\mathrm{d}\tau\frac{U}{2}n_{0}(\tau)[n_{0}(\tau)-1]
         +\int_{0}^{\beta}\mathrm{d}\tau\mathbf{\Phi}_{0}^{\dag}\mathbf{b}_{0}(\tau).
\end{eqnarray}
 In this equation, $\mathbf{\Phi}_{0}$ is related to the
condensation via
\begin{eqnarray}\label{eq:5}
\mathbf{\Phi}_{0}^{\dag}=\left(\begin{array}{cc}
                         -\tilde{t}\langle b_{0}^{\dag}\rangle_{S_{eff}}
                         & -\tilde{t}\langle b_{0}\rangle_{S_{eff}} \\
                         \end{array}\right).
\end{eqnarray}
Here $\tilde{t}$ is the hopping amplitude after the scaling has been
carried out. $\langle b_{0}^{\dag} \rangle$ is treated as a
$\tau$-independent quantity since we are studying an equilibrium
theory. The Weiss field $\mathbf{\mathcal{G}}_{0}^{-1}(i\omega_{n})$
represents the effective field from the environmental fluctuations
acting on the impurity site.

The self energy is defined through the Dyson equation
\begin{eqnarray}\label{eq:6}
\mathbf{\Sigma}(i\omega_{n})=2\mathbf{\mathcal{G}}_{0}^{-1}(i\omega_{n})-\mathbf{G}^{-1}_{c}(i\omega_{n}).
\end{eqnarray}
Here, $\mathbf{G}_{c}$ is the connected GF defined as
\begin{eqnarray}\label{eq:7}
\mathbf{G}_{c}(\tau-\tau')=\mathbf{G}(\tau-\tau')-\mathbf{G}_{dis}(\tau-\tau'),
\end{eqnarray}
where $\mathbf{G}_{dis}(\tau-\tau')$ is the disconnected part. Its
fourier transform is given in Appendix C. In the imaginary time axis
it is a constant and coincides with the condensed fraction in the
thermal dynamical limit. $\mathbf{\Sigma}$ and
$\mathbf{\mathcal{G}}_{0}$ have the same symmetry properties as the
GF. Among the four matrix elements only two functions are
independent. It is noted that the definition of the self-energy in
Eq.(\ref{eq:6}) has a factor of 2 difference from Eq.(11) in
Ref.~\onlinecite{BV}. This difference can be traced back to the
different conventions used for path integrals in Nambu
representation.\cite{DV} We have checked that our equations are
self-consistent and they guarantee $\mathbf{\Sigma}(i\omega_{n})=0$
for $U=0$.

The connected local GF of the lattice Hamiltonian is given by the
lattice Dyson equation
\begin{eqnarray}\label{eq:8}
\mathbf{G}_{c}(i\omega_{n})=\frac{1}{N_{latt}}\sum_{k}\left[[\mathbf{G}_{c}^{(0)}]^{-1}(k,i\omega_{n})
                        -\mathbf{\Sigma}(i\omega_{n})\right]^{-1}.
\end{eqnarray}
Here $N_{latt}$ is the total lattice number. $\mathbf{G}_{c}^{(0)}(k,i\omega_{n})$ is the connected GF of
the non-interacting system
$H_{0}=\sum_{k}(\epsilon_{k}-\mu)b_{k}^{\dagger}b_{k}$. It reads
\begin{eqnarray}\label{eq:9}
\mathbf{G}_{c}^{(0)}(k,i\omega_{n})=\left[i\omega_{n}\bold{\sigma_{3}}-(\epsilon_{k}-\mu)\mathbf{I}\right]^{-1}.
\end{eqnarray}
In the actual calculations, we transform the summation over $k$ in
Eq.(\ref{eq:8}) into the integral over energy. The explicit integral
formulas involving the semicircular density of states
Eq.(\ref{eq:2}) are summarized in Appendix A.
Eq.(\ref{eq:3})-Eq.(\ref{eq:9}) constitute the B-DMFT
self-consistency equations for the BHM.

\subsection{Impurity Solver}

In order to solve the B-DMFT equations, a suitable impurity solver
should be selected. To avoid technical complexities we use the exact
diagonalization method to solve the impurity model. It is simple,
fast, while at the same time qualitatively keeps the nontrivial
many-body physics of the problem \cite{CK}. The effective impurity
Hamiltonian equivalent to the action Eq.(\ref{eq:4}) reads
\begin{eqnarray}\label{eq:10}
H_{imp}&=&\sum_{k=1}^{B_{s}}\mathbf{a}_{k}^{\dag}\mathbf{E}_{k}\mathbf{a}_{k}
          +\sum_{k=1}^{B_s}(\mathbf{a}_{k}^{\dag}\mathbf{V}_{k}\mathbf{b}_{0}
          +\mathbf{b}_{0}^{\dag}\mathbf{V}_{k}^{\dag}\mathbf{a}_{k})\nonumber\\
       & &+\frac{U}{2}n_{0}(n_{0}-1)+\mathbf{\Phi}_{0}^{\dag}\mathbf{b}_{0}.
\end{eqnarray}
The creation and annihilation operators $\mathbf{a}_{k}^{\dag}$ and
$\mathbf{a}_{k}$ are for the environmental degrees of freedom and
are all in the Nambu representation. $B_{s}$ is the number of bath
sites. $\mathbf{E}_{k}$ and $\mathbf{V}_{k}$ are the kinetic energy
of environmental bosons and the coupling strength between the
environment and the impurity, respectively. They are $2\times 2$
matrices,
$\mathbf{E}_{k}=\left(\begin{array}{cc}
                           E_{k1} & E_{k2} \\
                           E_{k3} & E_{k4} \\
                           \end{array}
                           \right)$
and
 $\mathbf{V}_{k}=\left(\begin{array}{cc}
                    V_{k1} & V_{k2} \\
                    V_{k3} & V_{k4} \\
                    \end{array}
                    \right)$.
From the Hermiticity of $H_{imp}=H^{\dag}_{imp}$, we have $E_{k4}=E_{k1}$ being real,
$E_{k3}=E^{\ast}_{k2}$, $V_{k3}=V^{\ast}_{k2}$ and $V_{k4}=V^{\ast}_{k1}$. The requirement
that $H_{imp}$ is equivalent to the effective action $S_{eff}$ in Eq.(\ref{eq:4}) gives the
following relation (see Appendix B) between $\mathbf{\mathcal{G}}_{0}^{-1}$ and
$\mathbf{E}_{k},\mathbf{V}_{k}$,
\begin{eqnarray}\label{eq:11}
&& \mathbf{\mathcal{G}}_{0}^{-1}(i\omega_n)  \nonumber\\
&=&\left[\frac{1}{2}i\omega_{n}\mathbf{\sigma}_{3}
  +\frac{1}{2}\mu\mathbf{I}-\sum_{k=1}^{B_{s}}\mathbf{V}_{k}
  \left(\frac{1}{2} i\omega_{n}\mathbf{\sigma}_{3}
  -\mathbf{E}_{k} \right)^{-1}\mathbf{V}_{k}^{\dag} \right].  \nonumber \\
&&
\end{eqnarray}

We solve the B-DMFT equations using an iterative scheme as usually done for fermions. We start
from an initialization of the parameters $\mathbf{E}_{k}$, $\mathbf{V}_{k}$ ($k=1,..,B_s$)
and $\mathbf{\Phi}_{0}$. With them we calculate $\mathbf{\mathcal{G}}_{0}^{-1}(i\omega_{n})$
and define the impurity model Eq.(\ref{eq:10}). The impurity Hamiltonian is then solved by ED
to produce the connected GF $\mathbf{G}_{c}$ and a new $\mathbf{\Phi}_{0}$, according to the
following equation,
\begin{eqnarray}\label{eq:12}
\mathbf{G}_{c}(i\omega_{n})=\mathbf{G}(i\omega_{n})-\mathbf{G}_{dis}(i\omega_{n}),
\end{eqnarray}
and
\begin{eqnarray}\label{eq:13}
\mathbf{\Phi}_{0}=-\tilde{t}\langle \mathbf{b}_{0}\rangle.
\end{eqnarray}
Here $\langle ... \rangle$ represents the average under $H_{imp}$. $\mathbf{G}$ is calculated
from the Lehmann representation. $\mathbf{G}_{dis}(i\omega_{n})$ is the disconnected Green's
function. Details are in Appendix C.

Using Eq.(\ref{eq:6}), one obtains the self-energy $\mathbf{\Sigma}$ from $\mathbf{G}_{c}$. It
is then put into the lattice Dyson equation Eq.(\ref{eq:8}) to produce a new $\mathbf{G}_{c}$.
Using Eq.(\ref{eq:6}) again, we update the Weiss field $\mathbf{\mathcal{G}}_{0}$ and from it
we get the new parameters $\mathbf{E}_{k}$ and $\mathbf{V}_{k}$ through the following fitting
procedure.\cite{GK} A distance function $D\left[ \mathbf{E}_{k}, \mathbf{V}_{k} \right]$ is
defined as
\begin{eqnarray}\label{eq:14}
&& D\left[ \mathbf{E}_{k}, \mathbf{V}_{k} \right]  \nonumber\\
&=&\sum_{n} \left[ \left|\mathbf{\mathcal{G}}_{01}^{-1}(i\omega_{n})
-\widetilde{\mathbf{\mathcal{G}}}_{01}^{-1}(i\omega_{n}) \right|
+\left|\mathbf{\mathcal{G}}_{02}^{-1}(i\omega_{n})
-\widetilde{\mathbf{\mathcal{G}}}_{02}^{-1}(i\omega_{n}) \right| \right].  \nonumber \\
&&
\end{eqnarray}
Here $\widetilde{\mathbf{\mathcal{G}}}_{0}^{-1}$ is calculated from
$\mathbf{E}_{k}$ and $\mathbf{V}_{k}$ through Eq.(\ref{eq:11}).
$D\left[ \mathbf{E}_{k}, \mathbf{V}_{k} \right]$ is then minimized
with respect to $\mathbf{E}_{k}$ and $\mathbf{V}_{k}$ to find the
optimal parameters that can reproduce
$\mathbf{\mathcal{G}}_{0}^{-1}(i\omega_n)$ best. With these optimal
parameters we define a new impurity model to be diagonalized again.
The iteration continues until the lattice GF converges.

One specialty of boson is that it has an infinitely large local
Hilbert space. This poses difficulty for ED-based numerical methods
when adapted for bosons.\cite{RB} In our ED calculations, we
truncate the local Hilbert space by using $N+1$ boson states for
each mode, with $N$ a finite number. As the simplest algorithm, we
use the boson number eigen state $|n\rangle$ ($n=0,1,...,N$) as our
local basis, and keep the implementation of optimal basis for a
future improvement.\cite{CZ} The truncation of boson Hilbert space
introduces additional approximation and influences the accuracy of
our results, especially in the BEC phase (see below). It is
therefore important to check our results with respect to $N$ and to
make sure that the truncation errors are under control.

However, the truncation described above introduces a new problem to
the commutation relation of boson operators. In the truncated
Hilbert space, one has
\begin{equation} \label{eq:15}
b=\left(\begin{array}{cccccc}
                 0 & 1 & 0 & \cdots & 0 & 0 \\
                 0 & 0 & \sqrt{2} & \cdots & 0 & 0 \\
                 0 & 0 & 0 & \cdots & 0 & 0 \\
                 \vdots & \vdots & \vdots & \cdots & \vdots & \vdots \\
                 0 & 0 & 0 & \cdots & 0 & \sqrt{N} \\
                 0 & 0 & 0 & \cdots & 0 &  0  \\
               \end{array}
             \right)\\,
\end{equation}
and $b^{\dagger}$ is the hermitian conjugate matrix of $b$. From these one gets
$bb^{\dagger}=diag\{1,2,...,N,0 \}$ and $b^{\dagger}b=diag\{0,1,...,N-1,N \}$. The commutation
relation reads $[b, b^{\dagger}]=diag\{1,1,...,-N \}$ with the incorrect trace $Tr [b,
b^{\dagger}]=0$. Therefore, using representation Eq.(\ref{eq:15}) in our calculation will lead
to incorrect weight in the GFs as well as in the density of states. This problem cannot be
remedied by increasing $N$. To overcome this difficulty, we modify the representation of $b$
($b^{\dagger}$ accordingly) into
\begin{equation} \label{eq:16}
b=\left(\begin{array}{cccccc}
                 0 & 1 & 0 & \cdots & 0 & 0 \\
                 0 & 0 & \sqrt{2} & \cdots & 0 & 0 \\
                 0 & 0 & 0 & \cdots & 0 & 0 \\
                 \vdots & \vdots & \vdots & \cdots & \vdots & \vdots \\
                 0 & 0 & 0 & \cdots & 0 & \sqrt{N} \\
                 0 & 0 & 0 & \cdots & 0 &  \sqrt{N+1}  \\
               \end{array}
             \right)\\.
\end{equation}
It produces $b^{\dagger}b=diag\{0,1,...,N-1,2N+1 \}$, and $[
bb^{\dagger} ]_{i,i} =i$ $(i=1,...,N+1)$ and $[bb^{\dagger}
]_{N,N+1}=[bb^{\dagger}]_{N+1,N}=\sqrt{N(N+1)}$. The trace $Tr [b,
b^{\dagger}]=0$ is still incorrect. However, $b^{\dagger}b$ from
representation Eq.(\ref{eq:15}) and $bb^{\dagger}$ from
Eq.(\ref{eq:16}), if combined together, give the correct trace $Tr
[b, b^{\dagger}]=N+1$ Therefore, our strategy is that for any
operators involving $b^{\dagger}b$, such as $\langle i|b^{\dagger}|j
\rangle \langle j|b|i \rangle$ in the Lehmann representation of the
diagonal GF, we use Eq.(\ref{eq:15}). For operators involving
$bb^{\dagger}$ such as $\langle i|b|j \rangle \langle
j|b^{\dagger}|i \rangle$, we use Eq.(\ref{eq:16}) (see Eq.(C2) and
(C3) in Appendix C). In this way, the truncation introduced boson
commutation problem is solved.

\begin{figure}[b!]
\begin{center}
\vspace{0.15in}
\includegraphics[clip,width=9cm,height=7cm]{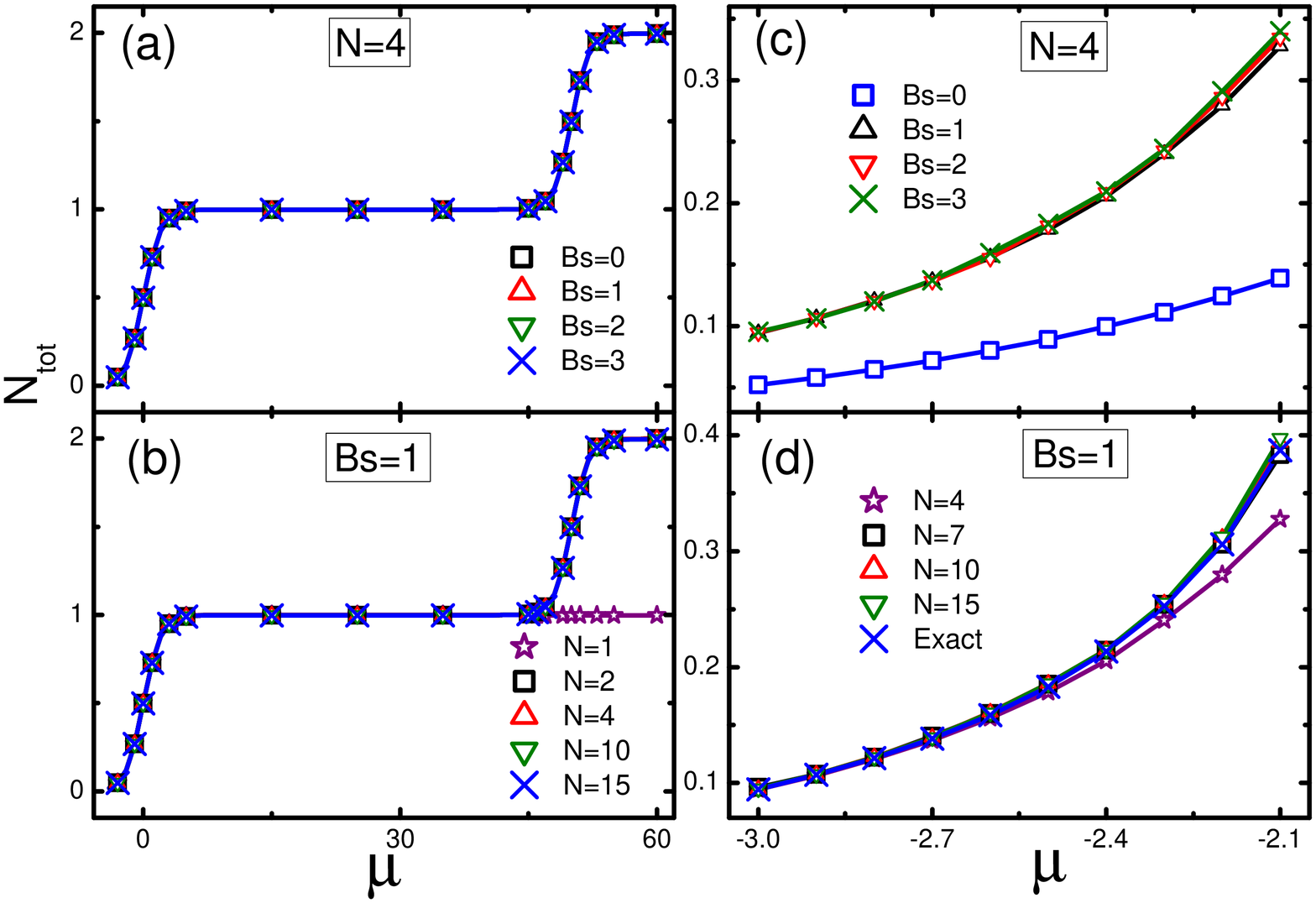}
\vspace*{-6pt}
\end{center}
\caption{(Color online) The total boson occupation as functions of chemical potential $\mu$.
(a) and (b): $U=50.0$, $\tilde{t}=0$, and $T=1.0$; (c) and (d): $U=0$, $\tilde{t}=1.0$, and
$T=1.0$.}
 \label{fig:1}
\end{figure}
For the bosonic impurity model Eq.(\ref{eq:10}) with $B_s$ bath
sites and $N+1$ states for each boson mode, the dimension of the
Hilbert space is $S_{t}=(N+1)^{B_{s}+1}$. To describe the BEC phase
where $\mathbf{\Phi}_{0}\neq 0$, the total particle number can no
longer be used as a good quantum number. In this case both ED and
calculating the GFs are very time consuming. As a result, the
parameters $N$ and $B_s$ are severely limited by the present
computer power. In the DMFT (ED) study of the Fermi Hubbard model
with four states on each site, it was shown that results converge
quite fast with the bath site number $B_{s}$, and $B_{s}=4$ already
gives qualitatively reliable results.\cite{GK} To explore the
$B_{s}$ and $N$ dependence of calculations for the Bose Hubbard
model, we calculate the $N_{tot}-\mu$ curves in both the atomic and
the free boson limit for different $N$ and $B_{s}$ values. The
results are shown in Fig.\ref{fig:1}. In both limits, as long as the
bath site number $B_{s}$ is larger than zero, the results are
already very close to the exact ones. At the same time,
$N$-dependence is more severe. The curves keep improving observably
until $N\geq10$. Taking a compromise between $N$ and $B_s$, in our
study we do all the calculations at $B_{s}=1$ (one bath site) and
$N=15$ (16 boson states) unless stated otherwise. We check our
results using larger $N$ and $B_s$ and make sure that our conclusion
does not depend on the selection of $B_{s}$ and $N$. For the
Matsubara frequencies $\omega_{n}=2n\pi/\beta$ in the GFs, we use
the cut off $|\omega_{n}|\leq 2000$.
\subsection{Non-Interacting Limit and Atomic Limit}
In this section, we check the B-DMFT formulas and our numerical results in the noninteracting
as well as in the atomic limit. For a noninteracting boson system, the bosons can move freely
in the lattices. They condense into a single particle state when the temperature is lower than
$T_{c}$. The exact solution of the Bose Hubbard model in this limit gives the thermally
excited boson occupation $N_{e}$ as
\begin{eqnarray}\label{eq:17}
N_{e}=\int_{-\infty}^{\infty}\mathrm{d}\epsilon
\frac{D(\epsilon)}{e^{\beta(\epsilon-\mu)}-1}.
\end{eqnarray}
The B-DMFT equations Eq.(\ref{eq:3})-Eq.(\ref{eq:9}) can also be
solved exactly in this limit.\cite{BV} At $U=0$, by carrying out the
Gaussian integral in Eq.(\ref{eq:4}) and doing the functional
derivative of the free energy with respect to
$\mathbf{\mathcal{G}}_{0}^{-1}$ (subtracting the disconnected
contribution), we get the connected GF as
\begin{equation} \label{eq:18}
\mathbf{G}_{c}(i\omega_{n})=\frac{1}{2}\mathbf{\mathcal{G}}_{0}(i\omega_{n})
\\
\end{equation}
which is independent of $\mathbf{\Phi}_{0}$. This gives a zero self-energy
$\mathbf{\Sigma}(i\omega_{n})=0$ according to Eq.(\ref{eq:6}), as expected. From
Eq.(\ref{eq:8}) and (\ref{eq:9}) we obtain the GF
\begin{equation}  \label{eq:19}
\mathbf{G}_{c}(i\omega_{n})=\int_{-\infty}^{\infty}\mathrm{d}\epsilon D(\epsilon)\left[
i\omega_{n}\mathbf{\sigma}_{3}-(\epsilon-\mu)\mathbf{I}\right]^{-1}.
\end{equation}
It is the exact result for the free bosons. The expression
Eq.(\ref{eq:17}) for the thermal excited particle number $N_{e}$ can
be recovered from it using the fluctuation-dissipation theorem.

The order parameter of BEC reads
 \begin{equation} \label{eq:20}
\begin{split}
\langle\mathbf{b}_{0}^{\dag}\rangle&=\frac{1}{Z}\int\prod_{i}\mathscr{D}b_{i}^{\ast}\mathscr{D}b_{i}
\mathbf{b}_{0}^{\dag}\exp(-S_{eff}) \\
 &=\frac{1}{2}\mathbf{\Phi}_{0}^{\dag}\mathbf{\mathcal{G}}_{0}(i0).
 \end{split}
 \end{equation}
Together with Eq.(\ref{eq:5}), it gives
\makeatletter
\let\@@@alph\@alph
\def\@alph#1{\ifcase#1\or \or $'$\or $''$\fi}\makeatother
\begin{subnumcases}
{\mathbf{\Phi_{0}}^{\dag}}
=0, &$\mu<-2\tilde{t}$, for normal phase, \label{eq:211}\\
\neq0, &$\mu=-2\tilde{t}$, for BEC phase.\label{eq:212}
\end{subnumcases}
\makeatletter\let\@alph\@@@alph\makeatother
Considering that for $\mu=-2\tilde{t}$, $\mathbf{G}_{c}(i0)=-(1/\tilde{t})\mathbf{I}$, we
cannot determine the nonzero value of $\mathbf{\Phi}_0$ in the BEC phase solely from the
self-consistency equations. This corresponds to the fact that for free bosons, the condensed
fraction cannot be fixed without giving the total particle number $N_{tot}$. The results above
can also be obtained from the equation of motion of GFs, starting from the impurity
Hamiltonian Eq.(\ref{eq:10}) with $U=0$. From this approach one gets
\begin{eqnarray}\label{eq:22}
\mathbf{G}_{c}(i\omega_{n})=\left[i\omega_{n}\mathbf{\sigma_{3}}+\mu\mathbf{I}-4\sum_{k}\mathbf{V}_{k}
\left(i\omega_{n}\mathbf{\sigma}_{3}-2\mathbf{E}_{k}
\right)^{-1}\mathbf{V}_{k}^{\dag} \right]^{-1},\nonumber \\
\end{eqnarray}
and
\begin{eqnarray}\label{eq:23}
\langle\mathbf{b_{0}}^{\dag}\rangle=\mathbf{\Phi}_{0}^{\dag}
\left[\mu\mathbf{I}+2\sum_{k}\mathbf{V}_{k}\mathbf{E}_{k}^{-1}\mathbf{V}_{k}^{\dag}
\right]^{-1}.
\end{eqnarray}
Substituting the parameters $\mathbf{E}_{k}$ and $\mathbf{V}_{k}$
with the $\mathbf{\mathcal{G}}_{0}$ in Eq.(\ref{eq:11}), we can get
the same results as Eq.(\ref{eq:18}) and (\ref{eq:20}).

\begin{figure}[b!]
\begin{center}
\vspace{0.15in}
\includegraphics[clip,width=9cm,height=7cm]{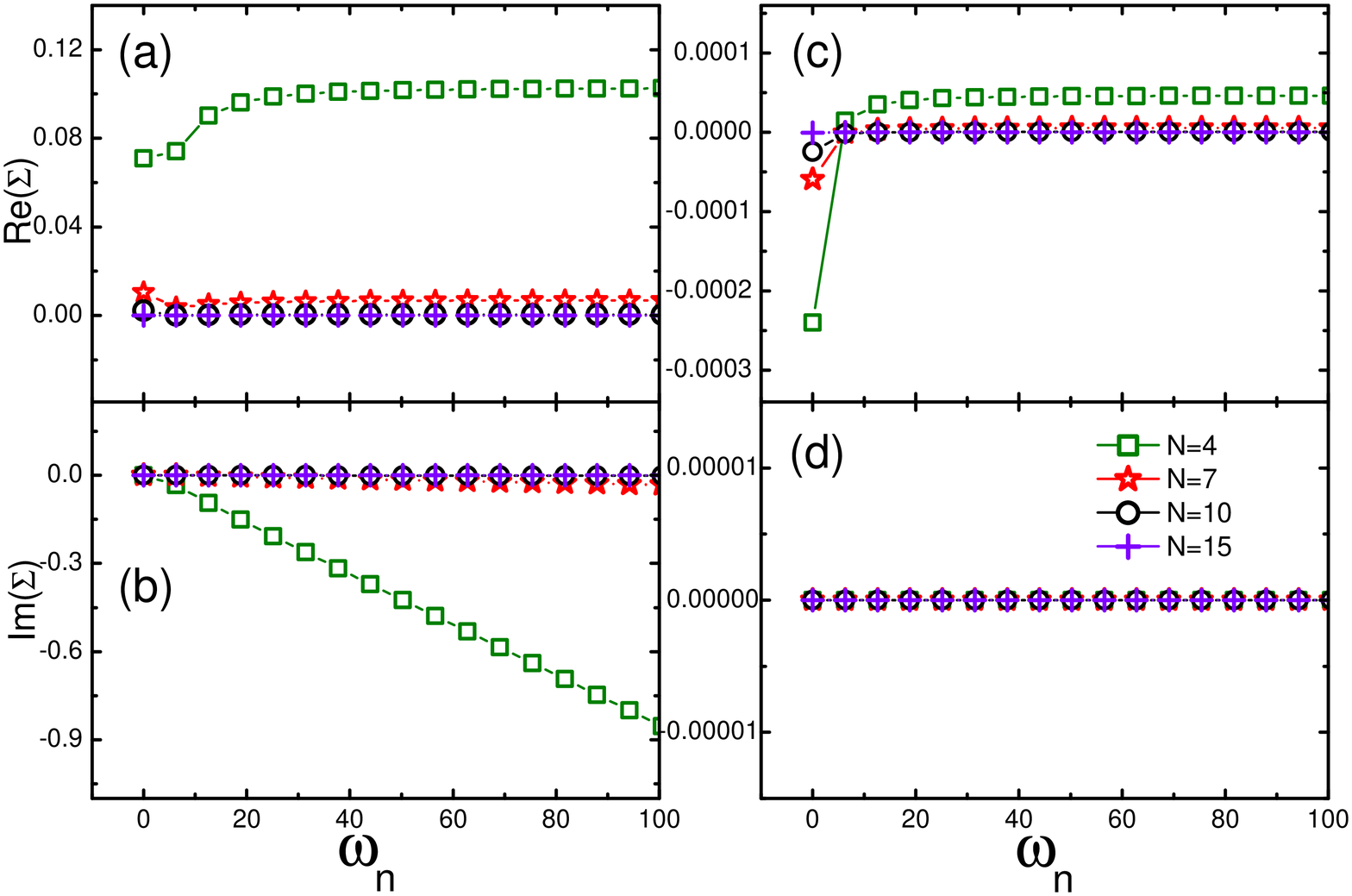}
\vspace*{-4pt}
\end{center}
\caption{(Color online) The self-energy of the non-interacting system as a function of
Matsubara frequency. (a) and (b): diagonal component $\mathbf{\Sigma}_{1}$; (c) and (d):
off-diagonal component $\mathbf{\Sigma}_{2}$. They are calculated at $U=0$, $\tilde{t}=1.0$,
$\mu=-2.1$, and $T=1.0$. Symbols are denoted in the figure. }
 \label{fig:2}
\end{figure}

\begin{figure}[b!]
\begin{center}
\vspace{0.15in}
\includegraphics[clip,width=9cm,height=6cm]{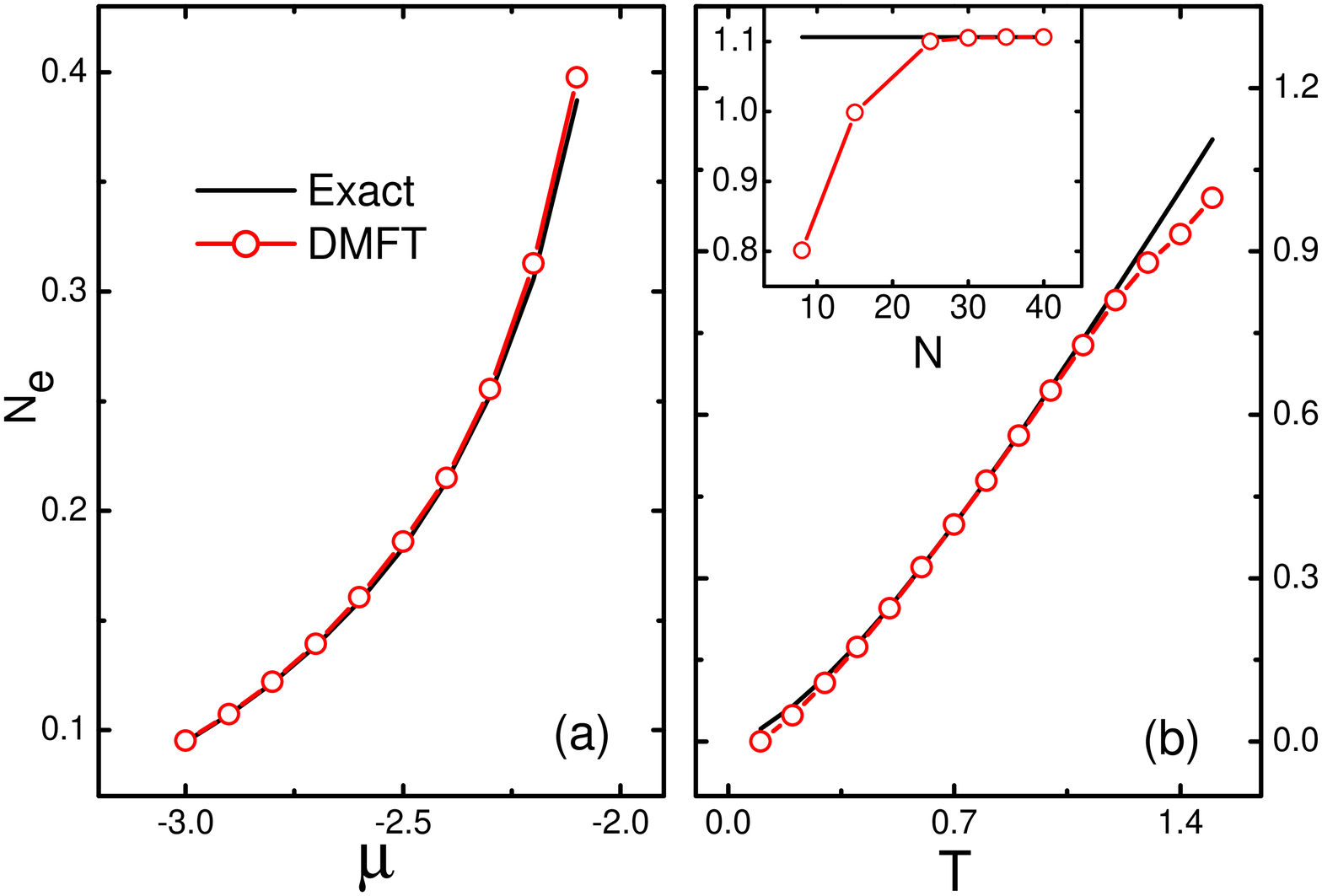}
\vspace*{-6pt}
\end{center}
\caption{(Color online) DMFT result (dots with eye-guiding lines) and the exact result (solid
line) for the thermal excited boson number $N_{e}$ at $U=0$, $\tilde{t}=1.0$. (a) as a
function of $\mu$ at $T=1.0$; (b) as a function of $T$ at $\mu=-2\tilde{t}$. Insert: $N_{e}$
as a function of $N$ at $\mu=-2\tilde{t}$, $T=1.5$.}
 \label{fig:3}
\end{figure}

In Fig.\ref{fig:2}, the real and imaginary part of the self-energy
are shown for $U=0$. Both the diagonal and off-diagonal components
tend to zero as the number of boson states $N$ increases. The self
energy is not strictly zero for finite $N$, because in the truncated
space, boson operators do not obey canonical commutation relations
and even $U=0$ does not correspond to a free system. From
Fig.\ref{fig:3}, it is seen that the $N_{e}-\mu$ curves from B-DMFT
at $U=0$ agree well with the exact ones. The deviation at high
temperatures decreases as $N$ increases, consistent with what we
find in Fig.\ref{fig:2}.

\begin{figure}[b!]
\begin{center}
\vspace{0.15in}
\includegraphics[clip,width=9cm,height=6cm]{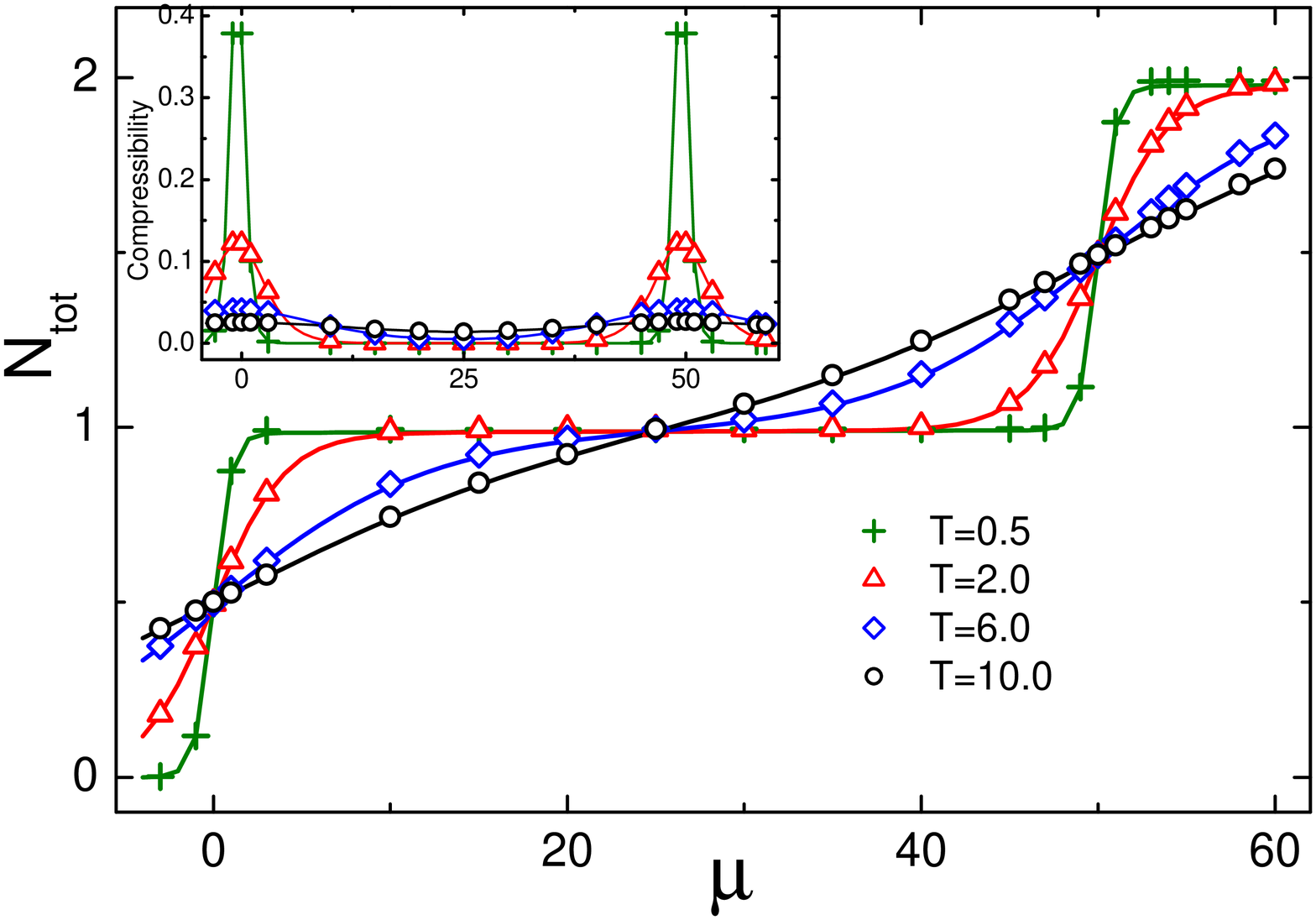}
\vspace*{-6pt}
\end{center}
\caption{(Color online) The total number of bosons as functions of the chemical potential
$\mu$ in the atomic limit with $U=50.0$ for different temperatures. The lines are the exact
results and symbols for B-DMFT results. Inset: the compressibility $\partial N_{tot}/\partial
\mu$ as functions of $\mu$ obtained from B-DMFT.}
 \label{fig:4}
\end{figure}
In the atomic limit $\tilde{t}=0$,  the quantum fluctuations of the
boson number operators disappear. Each site has an integer number of
localized particles at zero temperature. As the temperature
increases, thermal fluctuations dominate and the localized state
will melt gradually. In this limit, $\mathbf{\Phi}_{0}=0$ according
to Eq.(\ref{eq:5}). The density of states $D(\epsilon)$ becomes a
delta function, and Eq.(\ref{eq:8}) reduces to
\begin{eqnarray}\label{eq:24}
\mathbf{G}_{c}(i\omega_{n})=\left[i\omega_{n}\mathbf{\sigma}_{3}+\mu
\mathbf{I}-\mathbf{\Sigma}(i\omega_{n})\right]^{-1}.
\end{eqnarray}
Comparing with Eq.(\ref{eq:6}), one gets
$\mathbf{\mathcal{G}}_{0}^{-1}(i\omega_{n})=\left(i\omega_{n}
\sigma_{3}+\mu \mathbf{I} \right)/2$. When inserted into the
effective action Eq.(\ref{eq:4}), it gives exactly the action in the
atomic limit. Our numerical results for the atomic limit obtained
using $B_s=1$ and $N=15$ are shown in Fig.\ref{fig:4}. The B-DMFT
results (squares with guiding lines) agree well with the exact ones
(curves). Note that in the atomic limit, the B-DMFT results depend
very weakly on $B_{s}$ and $N$. It is seen that the thermal
activation will smear the Mott plateaus and the compressibility
$\partial N_{tot}/\partial\mu$ has broadened peaks. For $U=50$, the
Mott plateaus are clear at low temperatures and their features
disappear completely at about $T=10$. This observation agrees with
the conclusion that the MI melts completely at about $T^{*}=0.2U$ in
the limit $\tilde{t}=0$.\cite{G}
\section{Results and Discussions}
\begin{figure}[b!]
\begin{center}
\vspace{0.15in}
\includegraphics[clip,width=9cm,height=6cm]{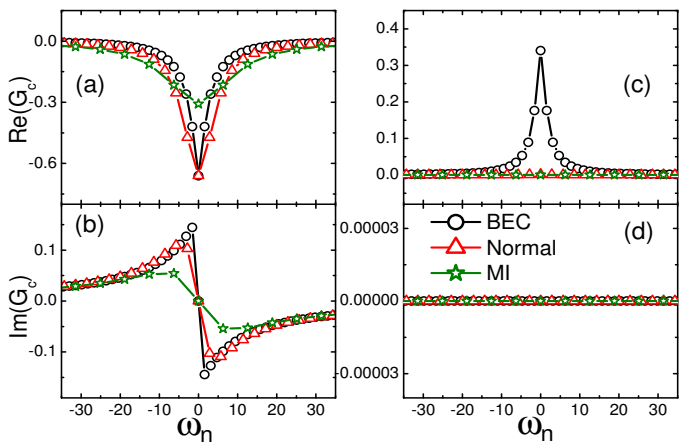}\\
\vspace*{-6pt}
\end{center}
\caption{(Color online) The connected diagonal GF $G_{c1}$ ((a) and
(b)) and off diagonal GF $G_{c2}$ ((c) and (d)) in the three phases:
BEC phase (circle): $\tilde{t}/U=0.2$; normal phase (triangle):
$\tilde{t}/U=0.11$; MI phase (pentacle): $\tilde{t}/U=0.05$. All are
calculated at $\tilde{t}=1.0$, $\mu/U=0.5$, and $T/U=0.05$. }
 \label{fig:5}
\end{figure}

\begin{figure}[b!]
\begin{center}
\vspace{0.15in}
\includegraphics[clip,width=9cm,height=6cm]{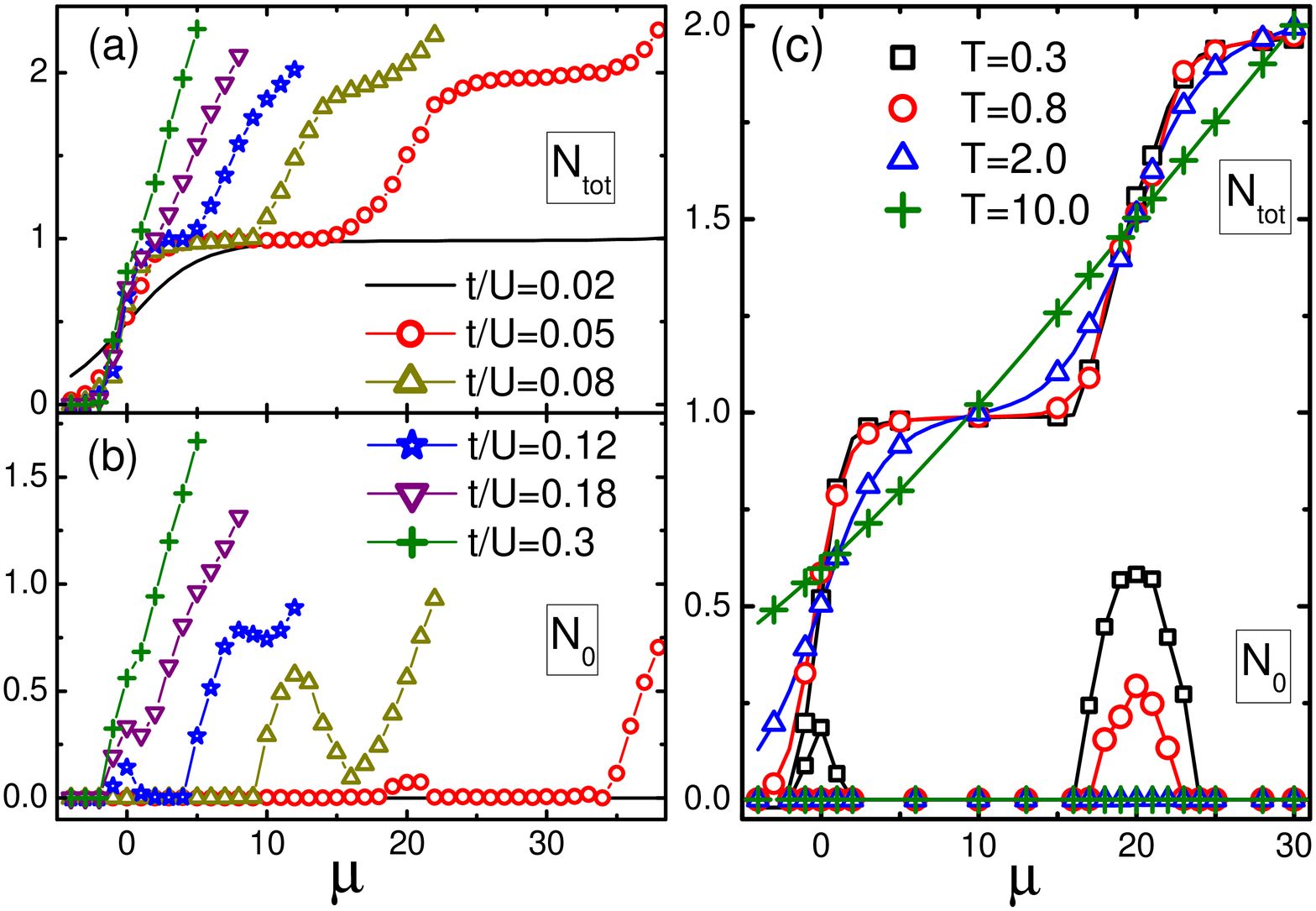}\\
\vspace*{-6pt}
\end{center}
\caption{(Color online) $N_{tot}$ and $N_{0}$ as functions of $\mu$.
(a) and (b) $T/U=0.05$ for different $\tilde{t}/U$
($\tilde{t}=1.0$); (c) $\tilde{t}=1.0$, $\tilde{t}/U=0.05$ for
different temperatures. Symbols with eye-guiding lines are denoted
in the figure. }
 \label{fig:6}
\end{figure}
In this section we discuss the physical results obtained by the
B-DMFT for the BHM. At zero temperature, the system should be either
in the BEC phase for weak interaction or in the MI phase for strong
interaction. At finite temperatures, besides the BEC and MI phases
that are extended from the ground state, there is the normal phase
that is connected to the MI and BEC phase in low temperature
regimes, through a crossover and a second-order phase transition,
respectively. Fig.\ref{fig:5}(a) and (b) show the diagonal component
of the converged connected GFs typical for the BEC, MI, and normal
phases. They are calculated at a finite but low temperature
$T/U=0.05$. All the high energy parts show the expected behavior
Re$G_{c1}(i\omega_n) \propto 1/\omega^{2}_{n}$ and
Im$G_{c1}(i\omega_n) \propto 1/\omega_{n}$. The low energy behaviors
are markedly different between the MI phase and the other two.
Similar to the MI phase of fermions, Im$G_{c1}(i\omega_n)
\rightarrow 0$ at zero frequency, signaling strong scattering and
nonexistence of well defined low energy quasiparticles in the MI
phase. In contrast, in the BEC and the normal phases, the diagonal
connected GFs are qualitatively similar. In Fig.\ref{fig:5}(c) and
(d) are the corresponding off diagonal GFs in the three phases. The
condensation in the BEC phase contributes to a sharp peak in the low
energy regime of Re$G_{c2}$. While in the MI and the normal phases,
no such peak appears. In all the three phases, Im$G_{c2}=0$ as it is
required by its definition and symmetry.

The full GF (not shown in Fig.\ref{fig:5}) can be written as the
form $G_{m}(i \omega_n)=G_{cm}(i \omega_n)+ \Delta_{m}
\delta_{n,0}$, $(m=1,2)$. For the BEC phase shown in
Fig.\ref{fig:5}, our numerical calculation gives $\Delta_{1}=-2.72$,
$\Delta_{2}=-2.71$, and $-\beta \langle b_{0}\rangle^2=-2.61$.
Within numerical errors, our result is consistent with the equation
$\Delta_{1}=\Delta_{2}=-\beta \langle b_{0}\rangle^2$ as can be seen
from the Lehmann expressions for GF in Appdix C. For the MI and the
normal phases, we always get $\Delta_1=\Delta_2=0$.

\begin{figure}[b!]
\begin{center}
\vspace{0.15in}
\includegraphics[clip,width=9cm,height=7cm]{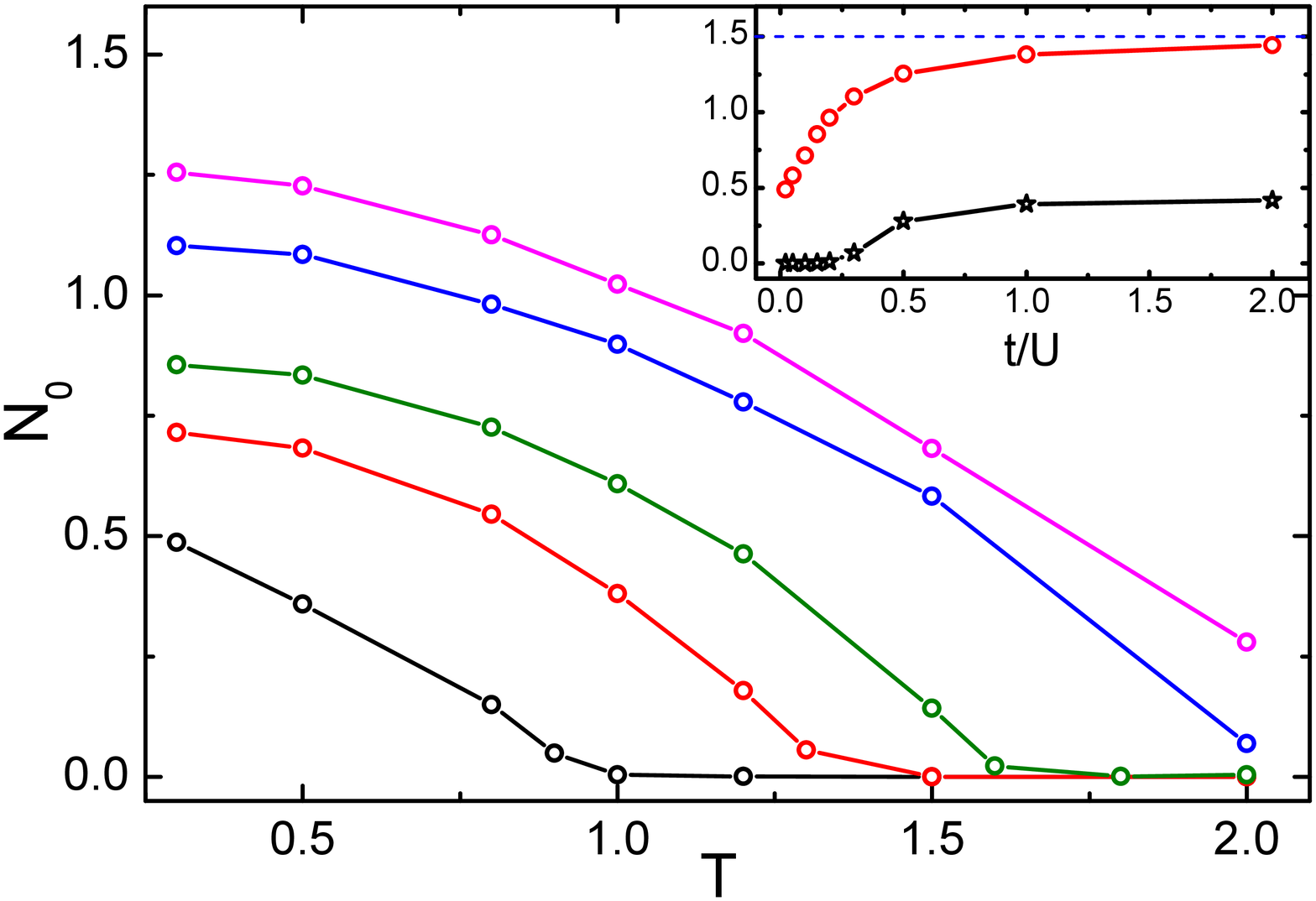}\\
\vspace*{-6pt}
\end{center}
\caption{(Color online) The condensed boson number $N_{0}$ as functions of temperature $T$ for
fixed total number $N_{tot}=1.5$ and different $\tilde{t}/U$ ($\tilde{t}=1.0$). From top to
bottom, $\tilde{t}/U=0.5, 0.3, 0.15, 0.1, 0.02$, respectively. Inset: $N_{0}$ changes with
$\tilde{t}/U$ for $T=0.3$ (circle) and $T=2.0$ (pentacle), respectively. The dashed line is
$N_{0}=1.5$. Solid lines are guiding lines.}
 \label{fig:7}
\end{figure}

\begin{figure}[b!]
\begin{center}
\vspace{0.15in}
\includegraphics[clip,width=9cm,height=7cm]{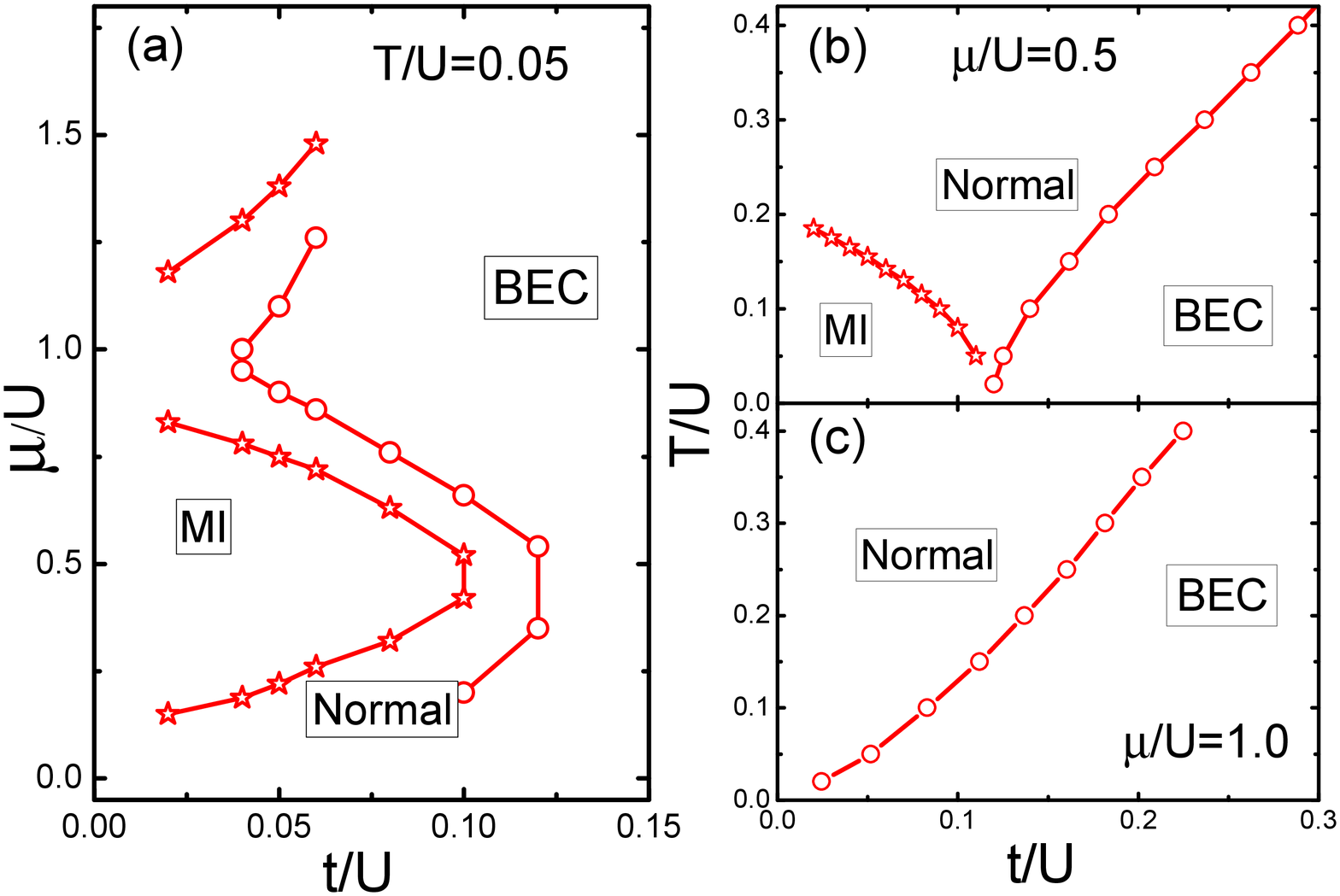}\\
\vspace*{-6pt}
\end{center}
\caption{(Color online) Phase diagrams. (a) in $\mu/U$ -
$\tilde{t}/U$ plane at $T/U=0.05$; (b) and (c) in $T/U$ -
$\tilde{t}/U$ plane at $\mu/U=0.5$ and $\mu/U=1.0$, respectively.
Three phases, BEC, normal phase, and the MI phase are marked out in
the figures. Lines are for eye-guiding.}
 \label{fig:8}
\end{figure}

 The total particle occupation is calculated by
\begin{eqnarray}\label{eq:25}
N_{tot}=-\frac{1}{\beta}\sum_{n}G_{c1}(i\omega_{n})e^{i\omega_{n}0^{+}}+\langle
b_{0}\rangle^{2},
\end{eqnarray}
where the condensed boson number $N_{0}$ is
\begin{eqnarray}\label{eq:26}
N_{0}=\langle b_{0}\rangle^{2}.
\end{eqnarray}
For $U>0$, a simple mean field analysis shows that the free energy becomes a quartic function
for large $\langle b_{0} \rangle$ and hence both $N_{0}$ and $N_{tot}$ can be determined
solely by the B-DMFT equations. They are plotted in Fig.\ref{fig:6} as functions of the
chemical potential. In Fig.\ref{fig:6}(a) we fix $T/U=0.05$ and study the evolution of the
curves as $\tilde{t}/U$ decreases. For large $\tilde{t}/U$ (small $U$ for fixed $\tilde{t}$),
$N_{tot}$ and $N_{0}$ are increasing functions of $\mu$ up to a boundary of $\mu$ and the
system always stay in the BEC phase. At the boundary, the convergence becomes slow and
difficult. As $\tilde{t}/U$ is smaller, a plateau of $N_{tot}=1$ begins to emerge in the
$N_{tot}-\mu$ curve. $N_{0}$ has a temporal decreases at the corresponding $\mu$ and then
continues to increase. The system is still in BEC phase, but the plateau and the dip in $N_0$
show the precursor to the MI phase. As $\tilde{t}/U$ still decreases, the plateau enlarges at
$N_{tot}=1$ and the next one at $N_{tot}=2$ begins to appear, forming Mott-like regimes.
$N_{0}$ has a well formed gap corresponding to each plateau, and has a peak signaling BEC
between two neighboring gaps. These BEC phases appear around $\mu=0, U, 2U, ...$ where two
adjacent Mott plateaus are connected. As $\tilde{t}/U$ decreases, the height of the $N_0$ peak
decreases and the critical temperature $T_{c}$ also decreases (see Fig.\ref{fig:8}(b)). For
very large U such as $\tilde{t}/U=0.02$, BEC doesn't appear any more because the critical
temperature is lower than the actual $T$, $(T/U)_{c} < T/U=0.05$ at $\tilde{t}/U=0.02$.
Fig.\ref{fig:6}(c) shows the temperature evolution at a fixed $\tilde{t}/U=0.05$. The
temperature effects on the Mott plateau as well as on the BEC phase are clearly observable. As
temperature rises, the Mott plateaus gradually blur at the shoulders and the condensed boson
number $N_{0}$ reduces to zero. For high enough temperature, the Mott plateaus finally
disappear and the MI crosses over to the normal phase.

A recent B-DMFT study for the bosonic Falicov-Kimball model reveals that the local repulsion
enhances the transition temperature of BEC.\cite{BV} Here we study the influence of $U$ on BEC
in the BHM. In Fig.\ref{fig:7}, we plot the $N_{0}-T$ curves at different $\tilde{t}/U$ values
for a fixed total density $N_{tot}=1.5$. For a given $\tilde{t}/U$, $N_{0}$ is a decreasing
function of $T$, and reduces to zero at $T_{c}$. With decreasing $\tilde{t}/U$ values
(increasing $U/\tilde{t}$), the $N_{0}-T$ curve shifts downwards, leading to smaller $N_{0}$
for a given $T$ and a reduction of $T_{c}$. This is consistent with the naive picture that
strong local repulsion between bosons tends to suppress the particle number fluctuations and
act against the BEC. Also, the quasiparticle states into which the bosons can condense are
turned into incoherent states and shifted into the Hubbard bands by a large $U$. Therefore,
our conclusion is that, different from the bosonic Falicov-Kimball model, the local repulsion
in the BHM reduces the transition temperature of BEC.

Different phases in the system can be distinguished from $N_{tot}$ and $N_{0}$. At zero
temperature, the system has two phases: the BEC phase and the MI. Due to the competition
between the on-site repulsion $U$ and the hopping $t$, there is a quantum phase transition
between them. As temperature increases, the BEC phase and the MI will change into normal phase
through a phase transition and a crossover, respectively. We have therefore three phases to
identify at finite temperatures: a nonzero $N_{0}$ signals the BEC phase, while the MI has
$N_{0}=0$ and an integer $N_{tot}$ with zero compressibility $\partial N_{tot}/\partial\mu$;
the phase with $N_{0}=0$ but a finite compressibility is the normal phase. According to this
criterion, we plot the phase diagrams in Fig.\ref{fig:8}.

In Fig.\ref{fig:8}(a) is the phase diagram on the
$\mu/U-\tilde{t}/U$ plane for finite temperature $T/U=0.05$. It is
obtained by scanning $\mu/U$ at fixed $\tilde{t}/U$. Three regimes
are clearly shown, the MI phase, BEC and the normal phase. Due to
the small $N$ parameter that we use, only the $N_{tot}=1$ and part
of the $N_{tot}=2$ MI domains are obtained. In the large
$\tilde{t}/U$ regime, BEC is stable. Between the two boundaries
(circles and pentacles) is the normal phase. The melting temperature
$T^{*}$ and BEC transition temperature $T_{c}$ are marked by
pentacles and circles, respectively. Similar finite temperature
phase diagram is also obtainable from a static mean field
theory\cite{BUV}.
 To understand the temperature effects on this diagram, we resort to
phase diagrams on the $T/U-\tilde{t}/U$ plane at two different
$\mu/U$ values, Fig.\ref{fig:8}(b) and (c). At $T=0$, a BEC-MI
quantum phase transition occurs at a critical $\tilde{t}/U$. The
difference between Fig.\ref{fig:8}(b) and (c) shows that
$(\tilde{t}/U)_c$ is dependent on $\mu/U$, consistent with the lobe
shape of the MI boundary in Fig.\ref{fig:8}(a). At finite
temperatures, the normal phase appears as a quantum critical regime
extending from the $T=0$ quantum critical point. This hints that in
the normal phase near $(\tilde{t}/U)_c$, critical behavior such as
power law correlation should exist. Experimental observation of such
quantum critical features in the normal phase near $(t/U)_c$ will be
an interesting issue.

A representative quantity for comparison between different theories
is the critical value $(t/U)_{c}$ at the tip of the the $n=1$
Mott lobe. It has been obtained by various methods.
For the 3D cubic lattice, the world line QMC gives
$(t/U)_{c}=0.032$(Ref. ~\onlinecite{KT}) and the worm algorithm
QMC gives $(t/U)_{c}=0.03408$ (Ref. ~\onlinecite{CSSS}). Recent
studies for the Bethe lattice with $z=6$ give $(t/U)_{c}=0.033$
(Ref. ~\onlinecite{SZ}) and $(t/U)_{c}=0.032$ (Ref.
~\onlinecite{HH}).  In our study, we obtained
$(\tilde{t}/U)_{c}=0.12$ for $\mu/U=0.5$ (n=1 Mott lobe). The
apparent discrepancy between our value and the previous ones is
because we didn't use the realistic lattice structure. In our
calculations, we take $z=\infty$ literally and set $\tilde{t}=1$ as
the energy unit, hence $z$ doesn't appear explicitly. Since
different scalings relating $\tilde{t}$ to $t$ are used in B-DMFT
for the normal and the condensed bosons, a critical value for $z=6$
cannot be simply recovered from our result by doing an inverse
scaling. A crude estimation, however, gives $(t/U)_c \approx 0.12/z
\sim 0.12/\sqrt{z}=0.02 \sim 0.049$ for $z=6$, being consistent with
the more accurate values.

For realistic lattices with a finite coordinate $z$, the B-DMFT is
still applicable but should be regarded as an approximation to
finite dimensional systems. For such a calculation, one should use
the actual dispersion $\epsilon_{k}$ of the given lattice in
Eq.(\ref{eq:9}), and replace $\tilde{t}$ with $zt$ in
Eq.(\ref{eq:5}). Experimentally, the BEC-MI transition point was
observed for $^{87}Rb$ atoms in 3D optical lattices.\cite{GB} The
transition occurs at a potential depth of $13E_{r}$ which compares
favorably with the mean-field value $U/t=5.8z$,\cite{OS,SR,BUV} but
differs from the more accurate QMC results cited above. The B-DMFT
calculation for the BHM in 3D cubic lattice and quantitative
comparison with the experiments as well as with the previous
theoretical results will be an interesting topic. But this is only
attainable when an accurate impurity solver is available. Therefore
we leave it for future study.

Finally we note that the ED method used in this work poses
limitations to our study. Due to finite number of boson states
$N=15$, reliable calculations are only possible in the small
$N_{tot}$ and small $N_{0}$ regimes. The small number of bath sites
$B_{s}=1$ causes slow convergence, especially near the MI-BEC
transition. Therefore, for practical applications of B-DMFT to boson
systems, it is necessary to develop an accurate and fast impurity
solver. In this respect, the recently developed bosonic NRG is a
promising technique.\cite{LB,RB} For the ED method, an algorithm
adopting the optimal boson basis will be interesting and progress is
being made in this direction.

\section{Conclusion}

In this paper we have performed the B-DMFT study for the BHM.
Following the ansatz of scaling in Ref.~\onlinecite{BV}, we obtain
the B-DMFT equations for the BHM. The bosonic effective impurity
Hamiltonian is solved by ED method with truncated boson Hilbert
space. We focus on the finite temperature properties of the
correlated bosons, and identify the MI, BEC and the normal phases.
The repulsive $U$ is found to suppress the BEC transition
temperature $T_{c}$. Phase diagrams on the $\mu/U-\tilde{t}/U$ and
$T/U-\tilde{t}/U$ planes are obtained, which disclose the quantum
critical nature of the low temperature normal phase. Relevance of
our results to other theoretical ones and the experimental
observations are discussed.

\begin{acknowledgments}
We thank Krzysztof Byczuk and Dieter Vollhardt for helpful
discussions and their comments on the manuscript. We also thank Anna
Kauch for pointing our an error in our formula. This work is
supported by NSFC under Grant No. 10674178 and the 973 program of
China (No. 2007CB925004).
\end{acknowledgments}
\begin{appendix}

\section{Integral of Semicircular Density of States}
The lattice Dyson equation in the B-DMFT equations (Eq.(\ref{eq:8})) is usually transformed
into an integral over $\epsilon$ of the form
\begin{eqnarray}\label{eq:A1}
 \mathbf{G}_{c}(i\omega_{n})&=&\int\mathrm{d}\epsilon D(\epsilon)[ i\omega_{n}\mathbf{\sigma}_{3}
 -(\epsilon-\mu)\mathbf{I}\nonumber\\
 & & -2\mathbf{\mathcal{G}}_{0}^{-1}(i\omega_{n})
 +[\mathbf{G}_{c}^{0}]^{-1}(i\omega_{n})]^{-1}.
\end{eqnarray}
One needs to carry out the integral for each $\omega_n$. For the
semicircular $D(\epsilon)$ given in Eq.(2)
\begin{eqnarray}\label{eq:2}
 D(\epsilon)=\frac{1}{2\pi t^{2}}\sqrt{4t^{2}-\epsilon^{2}}, &(|\epsilon|\leq
 2t),
\end{eqnarray}
the exact integral formula is given in the following.

\begin{widetext} \label{eq:A2}
\makeatletter
\let\@@@alph\@alph
\def\@alph#1{\ifcase#1\or \or $'$\or $''$\fi}\makeatother
\begin{subnumcases}
{\int_{-\infty}^{\infty}\mathrm{d}\epsilon
\frac{D(\epsilon)}{\xi-\epsilon}=}
\frac{\xi-\text{Sgn}\left(\text{Im}\xi \right)\sqrt{\xi^{2}-4t^{2}}}{2t^{2}}, & Im$\xi\neq0$, \label{eq:A21}\\
\frac{\xi-\text{Sgn}\left(\text{Re}\xi \right)\sqrt{\xi^{2}-4t^{2}}}{2t^{2}}, & Im$\xi=0$ and $|\xi|>2t$,\label{eq:B22}\\
\frac{\xi}{2t^{2}}, & Im$\xi=0$ and $|\xi|\leq2t$.\label{eq:A23}
\end{subnumcases}
\makeatletter\let\@alph\@@@alph\makeatother
\end{widetext}
In this equation, Im$\xi$ and Re$\xi$ are the imaginary and real parts of $\xi$. Sgn$(x)=1$
and Sgn$(x)=-1$ for $x$ being a positive and a negative real number, respectively.

\section{Effective Impurity Hamiltonian and Its Action}
The statistical action for the impurity model Eq.(\ref{eq:10}) reads
\begin{widetext}
\begin{eqnarray}\label{eq:B1}
 S_{imp}&=&\int_{0}^{\beta}\mathrm{d}\tau
           \left[\sum_{k=1}^{B_{s}}\mathbf{a}_{k}^{\dag}(\tau)\left(\frac{1}{2}\partial_{\tau}\mathbf{\sigma}_{3}
           +\mathbf{E}_{k} \right)\mathbf{a}_{k}(\tau)
           +\mathbf{b}_{0}^{\dag}(\tau) \left(\frac{1}{2}\partial_{\tau}\mathbf{\sigma}_{3}
           -\frac{1}{2}\mu I \right)\mathbf{b}_{0}(\tau)\right]\nonumber\\
        & &+\int_{0}^{\beta}\mathrm{d}\tau
           \left[\sum_{k=1}^{B_{s}} \left[\mathbf{a}_{k}^{\dag}(\tau)\mathbf{V}_{k}\mathbf{b}_{0}(\tau)
           +\mathbf{b}_{0}^{\dag}(\tau)\mathbf{V}_{k}\mathbf{a}_{k}(\tau) \right]
           +\frac{U}{2}n_{0}(\tau)[n_{0}(\tau)-1]
           +\mathbf{\Phi}_{0}^{\dag}(\tau)\mathbf{b}_{0}(\tau)\right].
\end{eqnarray}
\end{widetext}
The partition function can be expressed as the path integral over
complex boson fields

\begin{equation}\label{eq:B2}
   Z=\int\mathscr{D}b_{0}^{*}(\tau)\mathscr{D}b_{0}(\tau)
   \int\prod_{k=1}^{B_{s}}\mathscr{D}a_{k}^{*}(\tau)\mathscr{D}a_{k}(\tau)
   e^{-S_{imp}}.
\end{equation}
Carrying out the Gaussian integral for the environmental degrees of
freedom $\mathbf{a_{k}}^{\dag}$ and $\mathbf{a_{k}}$, one obtains
\begin{equation}\label{eq:B3}
   Z=Z_{a}\int\mathscr{D}b_{0}^{*}(\tau)\mathscr{D}b_{0}(\tau)
        e^{-S_{b}},
\end{equation}
where $Z_{a}$ is the partition function of the bath degrees of
freedom, and the effective action $S_{b}$ for the impurity is given
by

\begin{widetext}
\begin{eqnarray}\label{eq:B4}
 S_{b}=\int_{0}^{\beta}\mathrm{d}\tau
       \left[\mathbf{b}_{0}^{\ast}(\tau) \left[\frac{1}{2}\partial_{\tau}\mathbf{\sigma}_{3}-\frac{1}{2}\mu\mathbf{I}
       -\sum_{k=1}^{B_{s}}\mathbf{V}_{k}\left(\frac{1}{2}\partial_{\tau}\mathbf{\sigma}_{3}
       +\mathbf{E}_{k}\right)^{-1}\mathbf{V}_{k} \right]\mathbf{b}_{0}(\tau)
       +\frac{U}{2}n_{0}(\tau)[n_{0}(\tau)-1]
       +\mathbf{\Phi}_{0}^{\dag}(\tau)\mathbf{b}_{0}(\tau) \right]
\end{eqnarray}
\end{widetext}
Comparing this equation with the effective action derived from the
cavity method Eq.(\ref{eq:4}), one gets
\begin{eqnarray}\label{eq:B5}
&& \mathbf{\mathcal{G}}_{0}^{-1}(\tau-\tau')  \nonumber\\
&=&-\left[\frac{1}{2}\partial_{\tau}\mathbf{\sigma}_{3}
  -\frac{1}{2}\mu\mathbf{I}-\sum_{k=1}^{B_{s}}\mathbf{V}_{k}
  \left(\frac{1}{2}\partial_{\tau}\mathbf{\sigma}_{3}
  +\mathbf{E}_{k} \right)^{-1}\mathbf{V}_{k}^{\dag} \right]
  \delta(\tau-\tau').  \nonumber \\
&&
\end{eqnarray}
Through this equation the Weiss field
$\mathbf{\mathcal{G}}_{0}^{-1}$ is related to the impurity
parameters $\mathbf{E}_{k}$ and $\mathbf{V}_{k}$. After a Fourier
transform, one gets Eq.(\ref{eq:11}).

\section{Lehmann representation of the Boson Green's function}
The boson GFs are calculated from their Lehmann representation after the eigenvalues and the
eigenvectors are obtained by ED. In this appendix we present the corresponding Lehmann
representations. The GFs are defined in Eq.(\ref{eq:3}) as
\begin{equation} \label{eq:C1}
\begin{split}
\mathbf{G}(\tau-\tau')&\equiv -\langle T_{\tau}[\mathbf{b}(\tau)\mathbf{b}^{\dag}(\tau')]\rangle \\
                      &=\left(\begin{array}{cc}
                        -\langle T_{\tau}[b(\tau)b^{\dag}(\tau')]\rangle
                      & -\langle T_{\tau}[b(\tau)b(\tau')]\rangle \\
                        -\langle T_{\tau}[b^{\dag}(\tau)b^{\dag}(\tau')]\rangle
                      & -\langle T_{\tau}[b^{\dag}(\tau)b(\tau')]\rangle \\
                        \end{array}
                        \right) \\
                      &=\left(\begin{array}{cc}
                        G_{1}(\tau-\tau') & G_{2}(\tau-\tau') \\
                        G_{3}(\tau-\tau') & G_{4}(\tau-\tau') \\
                        \end{array}
                        \right).
                        \end{split}
\end{equation}
They have the symmetric relation
$G_{3}(\tau-\tau')=G_{2}^{\ast}(\tau-\tau')$ and
$G_{4}(\tau-\tau')=G_{1}^{\ast}(\tau-\tau')$. Here we only consider
$G_{1}$ and $G_{2}$.

The diagonal GF is expressed as
\begin{widetext}
for $\omega_{n}\neq0:$
\begin{eqnarray}\label{eq:C2}
G_{1}(i\omega_{n})&=&-\frac{1}{2Z}\sum_{ij}\frac{e^{-\beta
E_{j}}-e^{-\beta E_{i}}}{i\omega_{n}+(E_{i}-E_{j})} \langle
i|b|j\rangle\langle j|b^{\dag}|i\rangle
-\frac{1}{2Z}\sum_{ij}\frac{e^{-\beta E_{i}}-e^{-\beta
E_{j}}}{i\omega_{n}+(E_{j}-E_{i})} \langle
i|b^{\dag}|j\rangle\langle j|b|i\rangle;
\end{eqnarray}

for $\omega_{n}=0:$
\begin{eqnarray}\label{eq:C3}
G_{1}(i0)&=&-\frac{1}{2Z}\sum_{E_{i}\neq E_{j}}\frac{e^{-\beta
E_{j}}-e^{-\beta E_{i}}}{E_{i}-E_{j}} \langle i|b|j\rangle\langle
j|b^{\dag}|i\rangle -\frac{1}{2Z}\sum_{E_{i}\neq
E_{j}}\frac{e^{-\beta E_{i}}-e^{-\beta E_{j}}}{E_{j}-E_{i}} \langle
i|b^{\dag}|j\rangle\langle j|b|i\rangle\nonumber\\ &
&-\frac{\beta}{2Z}\sum_{E_{i}=E_{j}}e^{-\beta E_{i}}\langle
i|b|j\rangle\langle j|b^{\dag}|i\rangle
-\frac{\beta}{2Z}\sum_{E_{i}=E_{j}}e^{-\beta E_{i}}\langle
i|b^{\dag}|j\rangle\langle j|b|i\rangle.
\end{eqnarray}
\end{widetext}
Here, $Z=\sum_{i}exp(-\beta E_{i})$ is the partition function and
$\beta=1/T$ the inverse temperature. The eigenvectors can be
expanded by the basic vectors in the boson Fock space $\{ |n\rangle
\}$, $|i\rangle=\sum_{n} A^{i}_{n}|n\rangle$.
\addtocounter{equation}{1}
\begin{align}
\langle i|b|j\rangle&=\sum_{mn}A^{i*}_{n}A^{j}_{m}\langle n|b|m\rangle, \tag{\theequation a}\\
\langle j|b^{\dag}|i\rangle&=\sum_{mn}A^{j*}_{m}A^{i}_{n}\langle
m|b^{\dag}|n\rangle. \tag{\theequation b}
\end{align}
where $\langle n|b|m\rangle$ and $\langle m|b^{\dag}|n\rangle$ are
the matrix elements discussed in Eq.(\ref{eq:15}) and
Eq.(\ref{eq:16}). The coefficients $A^{i}_{n}$ can be obtained from
ED.

The off-diagonal GF reads
\begin{widetext}
for $\omega_{n}\neq0:$
\begin{eqnarray}\label{eq:C4}
G_{2}(i\omega_{n})&=&-\frac{1}{Z}\sum_{ij}\frac{e^{-\beta E_{j}}-e^{-\beta
E_{i}}}{i\omega_{n}+(E_{i}-E_{j})} \langle i|b|j\rangle\langle j|b|i\rangle,
\end{eqnarray}

for $\omega_{n}=0:$
\begin{eqnarray}\label{eq:C5}
G_{2}(i0)&=&-\frac{1}{Z}\sum_{E_{i}\neq E_{j}}\frac{e^{-\beta E_{j}}-e^{-\beta
E_{i}}}{E_{i}-E_{j}} \langle i|b|j\rangle\langle j|b|i\rangle
-\frac{\beta}{Z}\sum_{E_{i}=E_{j}}e^{-\beta E_{i}}\langle i|b|j\rangle\langle j|b|i\rangle.
\end{eqnarray}
\end{widetext}
The disconnected GFs $\mathbf{G}_{dis}(i\omega_{n})$ are defined as
the sum over $E_{i}= E_{j}$ parts in Eq.(\ref{eq:C3}) and
(\ref{eq:C5}).

\end{appendix}


\end{document}